\documentclass[article,twocolumn,preprintnumbers,superscriptaddress,nofootinbib]{revtex4-1}

\usepackage[dvipsnames]{xcolor}
\usepackage{enumerate}
\usepackage{mathrsfs,amsmath,amssymb}
\usepackage{natbib}
\usepackage{graphicx}
\usepackage{cancel}
\usepackage[normalem]{ulem}
\usepackage[pdftex,bookmarks,linktocpage,pdfpagelabels,plainpages=false,hyperfigures,linkcolor=blue,citecolor=blue]{hyperref} 
\hypersetup{colorlinks=true}

\newcommand\be{\begin{equation}}
\newcommand\ee{\end{equation}}
\newcommand\bea{\begin{eqnarray}}
\newcommand\eea{\end{eqnarray}}

\begin{document}
\bibliographystyle{apsrev4-1}

\title{Solar Gamma Ray Constraints on Dark Matter Annihilation to Secluded Mediators}

\author{Nicole F.~Bell}
\email{n.bell@unimelb.edu.au}
\affiliation{ARC Centre of Excellence for Dark Matter Particle Physics, School of Physics, The University of Melbourne, Victoria 3010, Australia}

\author{James B.~Dent} 
\email{jbdent@shsu.edu}
\affiliation{Department of Physics, Sam Houston State University, Huntsville, TX 77341, USA}

\author{Isaac W. Sanderson}
\email{isanderson@student.unimelb.edu.au}
\affiliation{ARC Centre of Excellence for Dark Matter Particle Physics, School of Physics, The University of Melbourne, Victoria 3010, Australia}

\begin{abstract}
We consider the indirect detection of dark matter that is captured in the Sun and subsequently annihilates to long-lived dark mediators. If these mediators escape the Sun before decaying, they can produce striking gamma ray signals, either via the decay of the mediators directly to photons, or via bremsstrahlung and hadronization of the mediator decay products.  Using recent measurements from the HAWC Observatory, we determine model-independent limits on heavy dark matter that are orders of magnitude more powerful than direct detection experiments, for both spin-dependent and spin-independent scattering. We also consider a well-motivated model in which fermionic dark matter annihilates to dark photons. For such a realistic scenario, the strength of the solar gamma ray constraints are reduced, compared to the idealistic case, because the dark matter capture cross section and mediator lifetime are related.  Nonetheless, solar gamma ray constraints enable us to exclude a previously unconstrained region of dark photon parameter space.
\end{abstract}

\maketitle

\section{Introduction}

The quest to reveal the true nature of dark matter (DM) is a forefront goal of modern science.  However, despite compelling cosmological evidence that allows us to infer the distribution of DM in the Universe, we have little information about the particles that constitute this matter, beyond the fact that they are not found within the standard model (SM), and their interactions with regular, SM matter must be very weak.  Currently, the most stringent constraints on such interactions are those set by the PICO-60~\cite{Amole:2019fdf} and XENON1T~\cite{Aprile_2019} direct detection experiments in the case of spin-dependent (SD) scattering of DM with protons and neutrons, respectively, and by XENON1T~\cite{Aprile_2018} in the case of spin-independent (SI) DM-nucleon scattering. In the face of null results from these increasingly sensitive experimental searches, it is important to move beyond overly simplistic assumptions about the nature of DM, such as the expectation of a generic weakly interacting massive particle (WIMP).  Moreover, it is important to explore complementary means of detection.

The capture of DM in astrophysical bodies such as the Sun
\cite{Press:1985ug,Griest:1986yu,Silk:1985ax,Krauss:1985ks,Gould:1987ju,Gould:1987ir,Jungman:1995df,Peter:2009mk,Zentner:2009is,Busoni:2013kaa,Garani:2017jcj,Busoni:2017mhe}, Earth~\cite{Gould:1987ww,Gould_2001,Green:2018qwo}, and other planets~\cite{Bramante:2019fhi,Garani:2019rcb,Leane:2020wob} or stars~\citep{Goldman:1989nd,Kouvaris:2007ay,Kouvaris:2010vv,deLavallaz:2010wp,McDermott:2011jp,Bell:2013xk,Bramante:2013nma,Bramante:2017xlb,Baryakhtar:2017dbj,Raj:2017wrv,Bell:2018pkk,Garani:2018kkd,Camargo:2019wou,Bell:2019pyc,Acevedo:2019agu,Joglekar:2019vzy,Joglekar:2020liw,Bell:2020jou,Ilie:2020vec,Dasgupta:2020dik,Bell:2020lmm,Bell:2020obw,Bell:2021fye}, provides an alternative means of probing DM scattering with regular matter. In the case of the Sun, DM-nucleon or DM-electron collisions can result in sufficient energy transfer for the DM to become gravitationally bound. Further energy loss, via subsequent collisions, causes the DM to accumulate in the solar core, where it may annihilate at a significant rate if the accumulated density becomes sufficiently large. Solar WIMP searches conventionally look for high energy neutrinos emanating from the solar core, as neutrinos are the only SM annihilation products that interact sufficiently weakly to escape the Sun. IceCube~\cite{Aartsen_2017b,Aartsen:2016exj} and Super-Kamiokande~\cite{Tanaka:2011uf,Choi:2015ara} place competitive constraints on SD DM-nucleon scattering cross sections, for annihilation channels that result in a hard neutrino spectrum. For other annihilation channels, and for SI scattering, these searches are not competitive with direct detection experiments. 
 
However, it is possible that DM annihilates not to SM states, but to dark sector particles which escape the Sun.  Indeed, most realistic DM models feature additional new particles, beyond the DM candidate itself. Particularly interesting, in the context of solar DM searches, are secluded models, in which the dark and visible sectors communicate only via a weakly coupled mediator~\cite{Pospelov:2007mp,Pospelov:2008jd,Schuster:2009fc,Schuster:2009au,Batell:2009zp,Rothstein:2009pm,Meade:2009mu,Bell:2011sn,Feng_2016,Feng:2016ijc,Leane:2017vag,Arina:2017sng,Leane:2021ihh}. Examples of such dark mediators include dark photons and many other well-motivated possibilities. The feeble strength of the mediator couplings is significant for two reasons: (i) the mediators will not be attenuated by scattering as they traverse the Sun; and (ii) they will be long-lived, and can propagate beyond the Sun before decaying back to SM states.  Annihilation of solar DM to such long-lived mediators therefore opens the possibility of dramatic new signals.

The decay of long-lived mediators, outside the Sun, can result in gamma rays~\cite{Batell:2009zp,Leane:2017vag,Arina:2017sng,Rothstein:2009pm,Leane:2021ihh}, charged particles~\cite{Feng_2016,Feng:2016ijc,Pospelov:2008jd,Meade:2009mu}, or enhanced (unattenuated) neutrino signals~\cite{Bell:2011sn,Adrian-Martinez:2016ujo,Ardid:2017lry,Niblaeus:2019gjk}. Our focus in this paper will be on gamma ray signals that result from the annihilation of heavy DM. In recent work, Leane et al.~\cite{Leane:2017vag} demonstrated that gamma ray data from Fermi-LAT~\cite{Abdo:2011xn,Ng:2015gya}
and HAWC~\cite{Albert:2018jwh,Albert:2018vcq} can be used to place powerful limits on DM annihilation to secluded mediators, in the case of a SD scattering cross section, surpassing direct detection constraints by many orders of magnitude.  In this paper, we will demonstrate that, surprisingly, solar gamma ray limits on SI interactions also surpass direct detection results, particularly for large DM mass.

We shall first calculate indicative model-independent limits for both SD and SI interactions.  In particular, following Ref.~\cite{Leane:2017vag} we shall assume that the DM-nucleon scattering cross section is energy independent, that the capture, annihilation, and decay rates are in general unrelated, and that the mediators decay between the Sun and the Earth.  While these model-independent limits are very strong,  they represent a simplistic and perhaps overly optimistic scenario.
In most realistic models, the DM capture and annihilation cross sections, and mediator lifetime, are not independent variables, but are instead determined by a common set of parameters. Therefore, we will also consider a model in which the interactions of fermionic DM are mediated by a dark photon that is kinetically mixed with the SM photon~\cite{Fabbrichesi:2020wbt,Kobzarev:1966qya,Holdom:1985ag,Holdom:1986eq,Feng:2015hja,Kouvaris:2016ltf}, allowing us to address a more realistic scenario.

In Section~\ref{sec:solarcap} we outline the calculation of the DM capture rate in the Sun, both for the generic case of an energy-independent SI or SD scattering cross section, and for the energy-dependent scattering cross section that arises in the context of the dark photon model. In Section~\ref{sec:annihilationrate} we outline annihilation rates for captured solar DM, as well as Sommerfeld enhancement of those annihilation rates for the dark photon model. In Section~\ref{sec:annihilationspec} we determine the gamma ray spectra that are generated via  decay of the
mediators, while in Section~\ref{sec:flux} we outline the calculation of the gamma ray flux and the comparison with HAWC data.
Our results are discussed in Section~\ref{sec:results}.

\section{Solar Capture of Dark Matter} 
\label{sec:solarcap}

\subsection{Capture in general}
The calculation of the capture and annihilation of DM in the Sun is well established
\cite{Press:1985ug,Griest:1986yu,Silk:1985ax,Krauss:1985ks,Gould:1987ju,Gould:1987ir,Jungman:1995df,Peter:2009mk,Zentner:2009is,Busoni:2013kaa,Garani:2017jcj,Busoni:2017mhe}. The number of DM particles in the Sun, $N_\chi$, follows the rate equation
\begin{align}
\label{eq:rate}    
\dot{N}_\chi = C_\text{cap} 
- C_\text{ann} N_\chi^2,
\end{align}
where $C_\text{cap}$ is the capture rate due to scattering on solar constituents and $C_\text{ann}N_\chi^2$ is the DM self-annihilation rate. 
For our model-independent analysis, we shall neglect the possibility of self-capture~\cite{Zentner:2009is} via scattering on accumulated DM. Self-capture is also  insignificant for the parameters of interest in the dark photon model we consider~\cite{Feng:2016ijc}. As we shall focus on TeV scale DM, we neglect the evaporation of DM from the Sun, which is negligible for DM masses above $\sim 4$~GeV \cite{Griest:1986yu,Gould:1987ju,Busoni:2013kaa,Busoni:2017mhe,Garani:2017jcj}.
We can then solve Eq.~(\ref{eq:rate}) to obtain
\begin{align}
    N_\chi &= \sqrt{\frac{C_\text{cap}}{C_\text{ann}}} \tanh{\frac{t}{\tau_\text{equil}}},
\end{align}
and hence,
\begin{align}
 \Gamma_\text{ann} &= \frac{1}{2} C_\text{ann} N_\chi^2 = \frac{1}{2} C_\text{cap} \tanh^2{\frac{t}{\tau_\text{equil}}}, 
 \label{eq:ann}
 \end{align}
where $\tau_\text{equil}$ is the time scale on which the capture and annihilation processes reach equilibrium,
\begin{align}
    \tau_\text{equil} = \frac{1}{\sqrt{C_\text{cap} C_\text{ann}}}.
\end{align}

The capture rate for scattering on a particular nuclear species, $N$, is given by 
\begin{align}
\label{eq:CNcap}    C^N_\text{cap} = n_\chi & \int^{R_\odot}_0 dr 4 \pi r^2 n_N(r) \int^\infty_0 dw 4 \pi w^3 f_\odot (w,r)   \nonumber \\
\times
&  \int dE_R \left. \frac{d \sigma_N}{d E_R} \right|_\text{capture},
\end{align}
where $w$ is the incident DM velocity, $E_R$ is the nuclear recoil energy, $n_\chi = \rho_\chi / m_\chi$ is the local DM number density, $n_N (r)$ is the number density of species $N$ at a distance $r$ from the solar center, $f_\odot (w,r)$ is the DM velocity distribution at that position, and $d\sigma_N / dE_R$ is the differential cross section for elastic DM-nuclei scattering. We take the solar density and element mass fractions from the AGSS09 model \cite{Serenelli_2009}.

The velocity, $u$, of DM asymptotically far from the Sun is assumed to follow a Maxwell-Boltzmann-like velocity distribution. In the neighbourhood of the Sun, this distribution is distorted due to the solar gravitational potential. Taking this acceleration into account and invoking energy conservation, the incoming DM velocity, $w$, for an interaction with a nucleus in the Sun is 
\begin{align}
\label{eq:velocity}    w^2 = u^2 + v_\odot^2 (r),
\end{align}
where $v_\odot (r)$ is the escape velocity at a distance $r$ from the solar center. We can thus express the integrand of Eq.~(\ref{eq:CNcap}) in terms of $u$ as
\begin{align}
    w^3 f_\odot (w,r) dw = u [u^2 + v_\odot^2 (r)] f_\odot(u) du,
\end{align}
where the asymptotic velocity distribution in the solar rest frame is
\begin{align}
    f_\odot (u) = \frac{1}{2} \int^1_{-1} dc f\left( \sqrt{u^2 + u_\odot^2 + 2 u u_\odot c} \right),
\end{align}
and $u_\odot = 233$ km/s is the solar velocity in the galactic rest frame. Numerical simulations suggest that the DM velocity  distribution function has the form~\cite{Vogelsberger_2009,Fairbairn_2009,Kuhlen_2010,Ling_2010,Lisanti_2011,Mao_2013}
\begin{align}
    f(u) = \tilde{N} \left[ \text{exp} \left( \frac{v_\text{gal}^2 - u^2}{k u_0^2} \right) -1 \right]^k \Theta (v_\text{gal} - u),
\end{align}
where $v_\text{gal}$ is the galactic escape velocity, $u_o$ is the velocity dispersion, $k$ is a power law index, and $\tilde{N}$ is a normalisation factor. A Maxwell-Boltzmann distribution is recovered for $k=0$ and $v_\text{gal} \to \infty$. The astrophysically favoured range of parameters is \citep{Baratella:2013fya}
\begin{align}
    220\;\text{km/s} < &u_0 < 270\;\text{km/s} \\
    450\;\text{km/s} < &v_\text{gal} < 650 \;\text{km/s} \\
    1.5 < &k < 3.5.
\end{align}

\bigskip

\subsection{Energy-independent capture}
For an energy independent DM-nucleus cross section $\sigma_N$, the contribution to the capture rate for a nuclear species $N$ becomes
\begin{align}
    C_\text{cap} = &n_\chi \sum_N \sigma_N \int^{R_\odot}_0 dr 4 \pi r^2 n_N(r) \nonumber \\
    &\times \int^\infty_0 du 4 \pi u [u^2 + v_\odot^2 (r)] f_\odot(u)   \mathcal{P}_N (u),
    \label{eq:Ccap}
\end{align}
where $\mathcal{P}_N (u)$ is the probability that a DM particle will be captured when scattering off a nucleus $N$.

For a SD DM-nucleon scattering interaction, the capture in the Sun is dominated by scattering from hydrogen~\cite{Baratella:2013fya} alone.  In this case, the nuclear form factor is $|F_N|^2 = 1$ for all relevant energies. Assuming the scattering cross section is energy independent, the capture probability can be simply approximated as~\cite{Baratella:2013fya}
\begin{align}
\mathcal{P}_N (u) &= \text{max} \left( 0, \frac{\Delta_\text{max} - \Delta_\text{min}}{\Delta_\text{max}} \right),
\label{eq:pn_approx}
\end{align}
where $\Delta_\text{max}=4 m_N m_\chi/(m_\chi + m_N)^2$ and $\Delta_\text{min}=u^2/(u^2 + u_{\odot \text{esc}}^2)$ are the maximal and minimal fractional energy losses that can result from scattering,
and $u_{\odot \text{esc}}$ is the escape velocity from the Sun.

In the case of SI scattering, we must account for the enhancement due to coherent scattering from all nucleons in the nucleus, and, in the case of heavier nuclei, a potentially significant suppression from the nuclear form factor. The probability of capture becomes
\begin{align}
    \mathcal{P}(u) = \frac{1}{E_{R\text{max}}} \int^{E_{R\text{max}}}_{E_{R\text{min}}} dE_R | F_N |^2,
    \label{eq:pn}
\end{align}
where $F_N$ is the Helmholtz form factor, given by
\begin{align}
    \left| F_N \right|^2 &= \text{exp} \left( -E_R / E_N \right),
\end{align}
with
\begin{align}
    E_N = \frac{5}{2m_i r_i^2} \approx  \frac{0.190 \text{ GeV}}{A_N^{5/3}}. 
\end{align}
One can scale the cross sections for heavier nuclei, $\sigma_N$, from that of the proton, $\sigma_p$~\cite{Baratella:2013fya}, as
\begin{align}
    \sigma_N &= \sigma_p \frac{m_N^2 (m_p+m_\chi)^2 A_N^2 }{m_p^2(m_N + m_\chi)^2},
\end{align}
where $m_p$ is the proton mass, $m_N$ is the nucleus mass, and $A_N$ is the number of nucleons in the nucleus. 

\subsection{Dark Photon model}

In order to go beyond model-independent limits, we shall consider a dark photon model as an example of a realistic benchmark scenario. We adopt a scenario in which fermionic DM couples to a dark photon mediator, which is the gauge boson associated with a broken $U(1)_{\text{dark}}$ gauge symmetry~\cite{Fabbrichesi:2020wbt}.  The kinetic mixing of  $U(1)_{\text{dark}}$ with the SM hypercharge results in mixing between the photon and the dark photon and, therefore, a small coupling of the dark photon to SM fermions. The effective Lagrangian is
\begin{align}
    \mathcal{L} \supset &-\frac{1}{4} F_{\mu \nu} F^{\mu \nu} - \frac{1}{4} F'_{\mu \nu} F'^{\mu \nu} + \frac{1}{2} m_{A'}^2 A'^2 \nonumber \\
    &- \sum_f q_f e (A_\mu + \varepsilon A'_\mu ) \bar{f} \gamma^\mu f - g_\chi A'_\mu \bar{\chi} \gamma^\mu \chi,
\end{align}
where $\varepsilon$ is the kinetic mixing parameter, $f$ are SM fermions of charge $q_f$, and $g_\chi$ is the $U(1)_{\text{dark}}$ gauge coupling. Such a dark photon may have evaded detection due to either a small mixing parameter, $\varepsilon$, and hence small coupling to SM particles, or a heavy mass~\cite{Fabbrichesi:2020wbt}. The case of a very small mixing parameter is of particular interest for solar gamma rays analyses, as the dark photon lifetime and scattering cross section will be such that they escape the Sun before decaying or scattering.

For DM-nucleus scattering mediated by the dark photon, the differential cross section with a nucleus $N$ in the non-relativistic limit \cite{Feng:2016ijc} is
\begin{align}
    \frac{d \sigma_N}{d E_R} &\approx 8 \pi \varepsilon^2 \alpha_\chi \alpha Z_N^2 \frac{m_N}{w^2 (2m_N E_R + m_{A'}^2)^2}|F_N|^2,
\end{align}
where $\alpha_\chi = g_\chi^2 / 4 \pi$ is the dark fine structure constant, $E_R$ is the recoil energy, and $Z_N$ is the number of protons in the nucleus. For the dark photon model, it can be shown that~\cite{Feng_2016}
\begin{align}
\int dE_R \left. \frac{d \sigma_N}{d E_R} \right|_\text{capture} = &\frac{2 \pi \varepsilon^2 \alpha_\chi \alpha Z_N^2}{w^2 m_N E_N} \exp \left( \frac{m_{A'}^2}{2m_N E_N} \right) \nonumber \\
&\times \left[ \frac{e^{-x_N}}{x_N} + \text{Ei} (-x_N) \right]^{x^{min}_N}_{x^{max}_N},
\end{align}
with 
\begin{align}
    x_N &= \frac{2m_N E_R + m_{A'}^2}{2m_N E_N}, \\
    \text{Ei}(z) &\equiv - \int^\infty_{-z}  \frac{e^{-t}}{t} dt,\\
    E_{R\text{min}} &= \frac{1}{2} m_\chi u^2, \\
    E_{R\text{max}} &= \frac{2 m_\chi^2 m_N}{(m_N + m_\chi)^2} \left( u^2 + v_\odot (r)^2\right),
\end{align}
where $E_\text{Rmin}$ and $E_\text{Rmax}$ define the minimum and maximum recoil energies for which capture occurs, respectively.
Combining the equations above with Eq.~(\ref{eq:CNcap}) results in
\begin{align}
    C^N_{\text{cap}} = 32 \pi^3 \varepsilon^2 \alpha_\chi \alpha n_\chi \frac{Z_N^2}{m_N E_N} \exp \left( \frac{m_{A'}^2}{2m_N E_N} \right) c^N_{\text{cap}} , 
\end{align}
where
\begin{align}
        c^N_\text{cap} & =   \int^{R_\odot}_0 dr r^2 n_N(r) \int^\infty_0 dw~w f_\odot (w,r), \nonumber \\
        &\times \left[ \frac{e^{-x_N}}{x_N} + \text{Ei} (-x_N) \right]^{x^{min}_N}_{x^{max}_N}.
\end{align}
Because the dark photon mediator couples to nuclei through kinetic mixing with the photon, the DM-nucleus cross section scales as $Z_N^2$, rather than the usual $A_N^2$ scaling that is applicable when DM couples to nucleons in an isospin invariant fashion.

\section{Annihilation Rate}

\label{sec:annihilationrate}

We shall assume that the DM annihilates to a pair of secluded mediators, $\chi\chi\rightarrow YY$, or, in the case of the dark photon model, $\chi\chi \rightarrow A'A'$. These dark mediators will then decay to produce fluxes of SM particles, including gamma rays.

The annihilation will occur in the centre of the Sun, after the DM has thermalised. Note that the thermalisation time will be much shorter than the age of the Sun for much of the parameter space of interest. However, the thermalisation assumption breaks down for very small cross sections or very large DM masses~\cite{Peter:2009mk}, significantly suppressing the annihilation rate. We shall indicate this region of parameter space in our results in Section~\ref{sec:results}.

Assuming the thermalisation assumption to be valid, the DM will form a small Boltzmann distributed core localised at the centre of the Sun, with number density
\begin{align}
    n_\chi = n_0 e^{-r^2 / r_\chi^2},
\end{align}
where
\begin{align}
    r_\chi = \sqrt{\frac{3T_\odot}{2\pi G_N \rho_\odot m_\chi}} 
    \approx 0.01 R_\odot \sqrt{\frac{100 \text{GeV}}{m_\chi}},
\end{align}
and a very small velocity
\begin{align}
    v_0 = \sqrt{2T_\odot / m_\chi} = 5.1 \times 10^{-5} \sqrt{\text{TeV} / m_\chi}.
\end{align}
For an annihilation rate of $\Gamma_\text{ann} = \frac{1}{2} C_\text{ann} N_\chi^2$, we have an annihilation coefficient of~\cite{Baratella:2013fya,Feng:2016ijc}
\begin{align}
    C_\text{ann} &= \langle \sigma_\text{ann} v \rangle \left( \frac{G_N m_\chi \rho_\odot}{3 T_\odot} \right)^{3/2},
\end{align}
with core solar density $\rho_\odot = 151$ g/cm$^3$ and temperature $T_\odot~=~15.5~\times10^6$ K.

Given the extremely slow velocities, Sommerfeld enhancement provides a significant boost to the annihilation rate in the dark photon scenario,
\begin{align}
    \langle \sigma_\text{ann} v \rangle &= S \langle \sigma_\text{ann} \rangle^\text{Born},
\end{align}
where $S$ is the non-relativistic Sommerfeld enhancement and $\langle \sigma_\text{ann} v \rangle^\text{Born}$ is the thermally averaged cross section in the Born approximation. For the dominant annihilation channel, $\chi \chi \to A' A'$, this is~\cite{Liu_2015}
\begin{align}
    \langle \sigma_\text{ann} v \rangle^\text{Born} &= \frac{\pi \alpha_\chi^2}{m_\chi^2} \frac{(1-m_{A'}^2/m_\chi^2)^{3/2}}{[1 - m_{A'}^2 / (2m_\chi^2)]^2}.
\end{align}
The thermally averaged enhancement to $s$-wave annihilation is given by~\cite{Cassel_2010,Feng_2010}
\begin{align}
    S &= \langle S_s \rangle = \int \frac{d^3 v}{(2 \pi v_0^2)^{3/2}} e^{-\frac{1}{2} v^2 / v_0^2} S_s, \\
    S_s &= \frac{\pi}{a} \frac{\sinh (2\pi a c)}{\cosh (2\pi a c) - \cos (2 \pi \sqrt{c-a^2 c^2})},
\end{align}
where $a = v/(2\alpha_\chi)$ and $c = 6\alpha_\chi m_\chi / (\pi^2 m_{A'})$.

If we fix the annihilation rate at freezeout to be $\langle \sigma_\text{ann} v \rangle = 2.2 \times 10^{-26} \text{ cm}^3/\text{s}$ to obtain the correct DM relic density, $\Omega_\chi h^2 = 0.12$, we require the dark fine structure constant to be
\begin{align}
    \alpha_\chi &= 0.024 \left( \frac{m_\chi}{\text{TeV}} \right).
\end{align}
Alternatively, we can fix $\alpha_\chi$ by setting it to the maximum value allowed by cosmic microwave background constraints~\cite{Slatyer_2016},
\begin{align}
    \alpha_\chi = 0.17 \left( \frac{m_\chi}{\text{TeV}} \right)^{1.61}.
\end{align}
In Section~\ref{sec:results}, we shall show results for $\alpha_\chi$ fixed through thermal freezeout and, for comparison, for a generic coupling of $g_\chi = \sqrt{4 \pi \alpha_\chi} = 1$.

\section{Annihilation Spectrum} 
\label{sec:annihilationspec}

\begin{figure}
    \centering
    \includegraphics[width = \columnwidth]{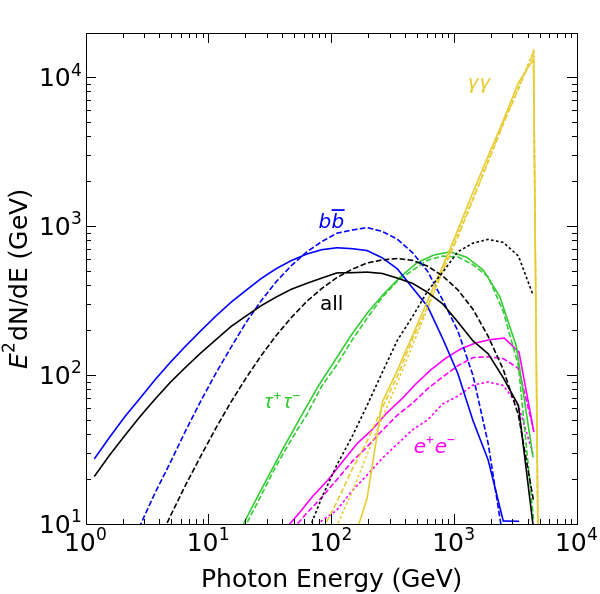}
    \caption{Comparison of spectrum per annihilation for $\chi\chi \to YY$ followed by $Y\rightarrow e^+ e^-$, $\bar{\tau} \tau$, $\bar{b} b$ and $\gamma \gamma$, for $m_Y = 1$ GeV (where kinematically accessible, dotted), $50$ GeV (dashed), and $2$ TeV (solid), using $m_\chi = 5$ TeV.}
    \label{fig:mediatorchannels}
\end{figure}

We shall assume the mediators, $Y$, decay either to a pair of SM fermions, or directly to a pair of photons. The charged fermion final states will result in a substantial flux of gamma rays, either via bremsstrahlung processes or via the decay of hadronic final states.
In the case of the dark photon model, the decay channel $A' \to \gamma \gamma$ is forbidden.  Instead, the $A'$ will decay to all kinematically available pairs of charged particles. While the charged particles could be probed with detectors such as AMS~\cite{Feng_2016,Feng:2016ijc,Aguilar:2013qda,Aguilar:2015ooa,Aguilar:2014mma}, the gamma ray fluxes can be probed by current experiments such as HAWC~\cite{Harding:2015bua} and future detectors such as LHASSO~\cite{Zha:2012wp} and SWGO~\cite{Abreu:2019ahw}.

We generate the spectrum of photons per annihilation using either Pythia8~\cite{Sjostrand:2006za,Sjostrand:2014zea} or analytically. In the case where the mediator mass falls in the range $0.05~{\rm{GeV}} ~\lesssim~m_Y~<~2~m_\mu$, this is well approximated by the analytic bremsstrahlung spectrum
\begin{align}
 \frac{dN_\gamma}{dx_1} &=  \int^{t_{1,\text{max}}}_{t_{1,\text{min}}} \frac{dx_0}{x_0 \sqrt{1 - \epsilon_1^2}} \frac{dN_\gamma}{dx_0}, \\
    \frac{dN_\gamma}{dx_0} &= \frac{\alpha_{\text{EM}}}{\pi} \frac{1 + (1-x_0)^2}{x_0} \left[ \ln \left( \frac{4 (1-x_0)}{\epsilon_f^2}\right) - 1 \right] + \mathcal{O} (\epsilon_f^2), \label{eq:brehm}
\end{align}
where
\begin{align}
        t_{1,\text{max}} &= \min \left[ 1, \frac{2x_1}{\epsilon_1^2} \left( 1 + \sqrt{1 - \epsilon_1^2} \right) \right], \\
    t_{1,\text{min}} &=  \frac{2x_1}{\epsilon_1^2} \left( 1 - \sqrt{1 - \epsilon_1^2} \right),
\end{align}
and $x_0 = 2 E_0 / m_Y$, where $E_0$ is the energy of a photon in the rest frame of the mediator $Y$,  $x_1 = E_1 / m_\chi$, $E_1$ is the photon energy in the center-of-mass frame of the $\chi \chi$ annihilation, $\epsilon_1= 2m_Y/m_\chi$, and $\epsilon_f = 2m_e / m_\chi$. Below $0.05$ GeV, the $\mathcal{O} (\epsilon_f^2)$ effects are no longer insignificant, and the approximation is no longer valid.

For the model-independent case of a mediator decay to a specific final state, the spectra per-annihilation is shown in Figure~\ref{fig:mediatorchannels} for various decay channels.
We see that the spectrum depends somewhat on the mediator mass. This is not just due to opening new channels -- the bremsstrahlung spectrum itself has approximately a logarithmic dependence on the mediator mass. 
For example, if we consider the  $\chi\chi \to YY \to e^+ e^-e^+e^-$ process, for the spectrum point $E_\gamma = m_\chi/2$, we find
\begin{align}
    \left. \frac{1}{m_\chi} \frac{dN_\gamma}{dE_\gamma} \right|_{E_\gamma = m_\chi/2}  &\simeq 1.11 \frac{\alpha}{\pi} \log \left[ \frac{m_Y^2}{4 m_f^2} \right]   - 1.06 \frac{\alpha}{\pi} \nonumber \\
    &+ \mathcal{O} \left( \frac{m_Y^2}{m_\chi^2} \right). 
\end{align}

\section{HAWC gamma ray constraints}
\label{sec:flux}

The latest HAWC observations of the Sun~\cite{Albert:2018vcq} provide constraints on the observed gamma ray flux from $\sim 1$~TeV to several hundred TeV. This is ideal for probing very heavy DM, for which the constraints from direct detection experiments are more modest.

For mediators which decay outside of the Sun, we calculate the expected energy flux as
\begin{align}
    E^2\frac{d\Phi}{dE} =\frac{\Gamma_\text{ann}}{4\pi D_\oplus^2}  E^2 \frac{dN}{dE} 
    \left( e^{-R_\odot/ \gamma c\tau} - e^{-D_\oplus/ \gamma c\tau}\right),
\end{align}
where $\Gamma_\text{ann}$ is the annihilation rate given by Eq.~(\ref{eq:ann}), 
$R_\odot$ is the radius of the Sun, $D_\oplus$ is the Earth-Sun distance, $\tau$ is the mediator lifetime, and $\gamma$ is the Lorentz factor of the mediator in the solar rest frame.
For our model-independent analysis,  we consider the optimistic case in which all mediators decay between the surface of the Sun and the Earth. This will be true to within a factor of 2 when the quantity $\gamma c \tau$ is in the range $R_\odot - D_\oplus$~\cite{Leane:2017vag}. We note that for $m_\chi \gg m_Y$, the photons will be collinear with the highly boosted mediator, hence we can treat the Sun as a gamma ray point source.

The HAWC observatory has reported 95\% C.L. upper limits on the gamma ray emission from the Sun, in five log-spaced energy bins spanning a range from $\sim 1$ TeV  to $\sim 100$ TeV~\cite{Albert:2018vcq}.  We consider our DM model parameters to be excluded if the expected gamma ray flux from mediator decays exceeds the HAWC limit for any energy bin.
For the decay channels that result in hard gamma ray spectra (particularly $\gamma\gamma$, $e^+e^-$, and $\tau^+\tau^-$) the limit will typically be set by the energy bin closest to $E_\gamma = m_\chi$. This is a conservative approach. Note that the HAWC data points feature very little scatter, which gives confidence that there is no problem with any individual data point -- see Fig.6 of Ref.~\cite{Albert:2018vcq}.  Therefore, a more detailed statistical analysis is expected to produce a stronger limit.

\begin{figure*}[t]
\includegraphics[width=0.45\textwidth]{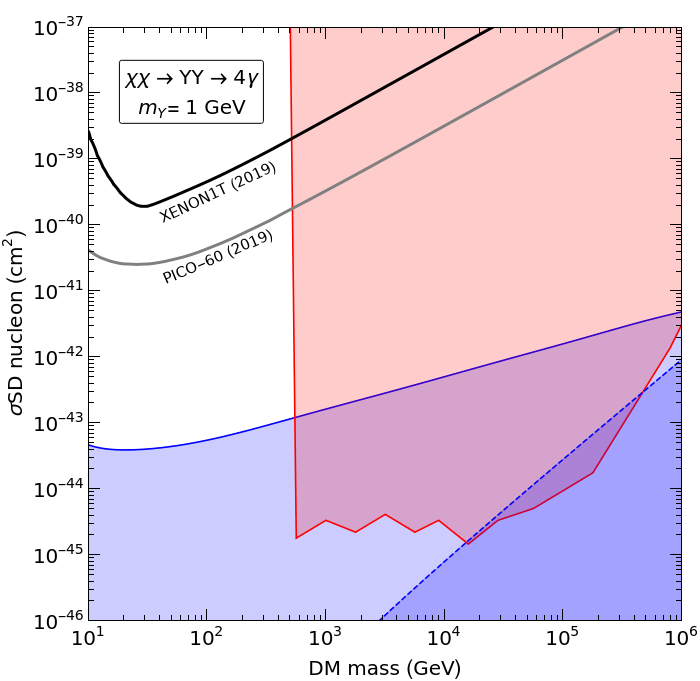} \includegraphics[width=0.45\textwidth]{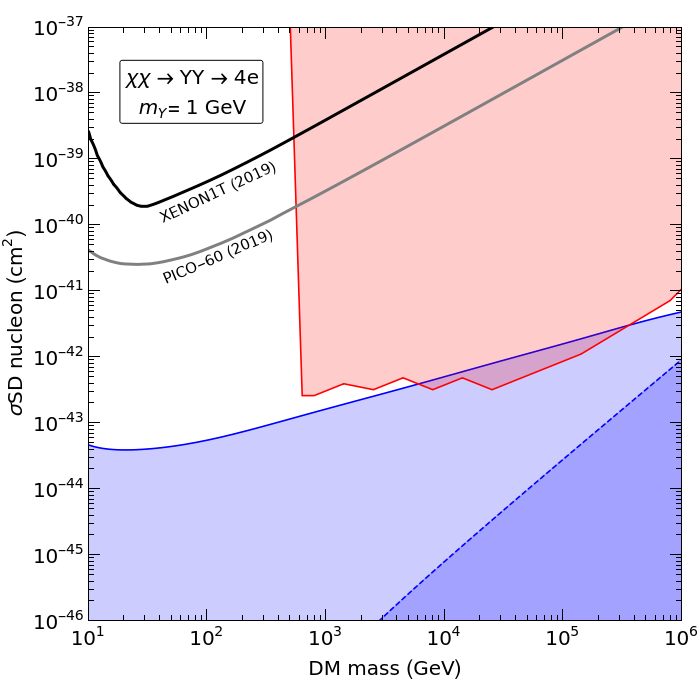} \\
\includegraphics[width=0.45\textwidth]{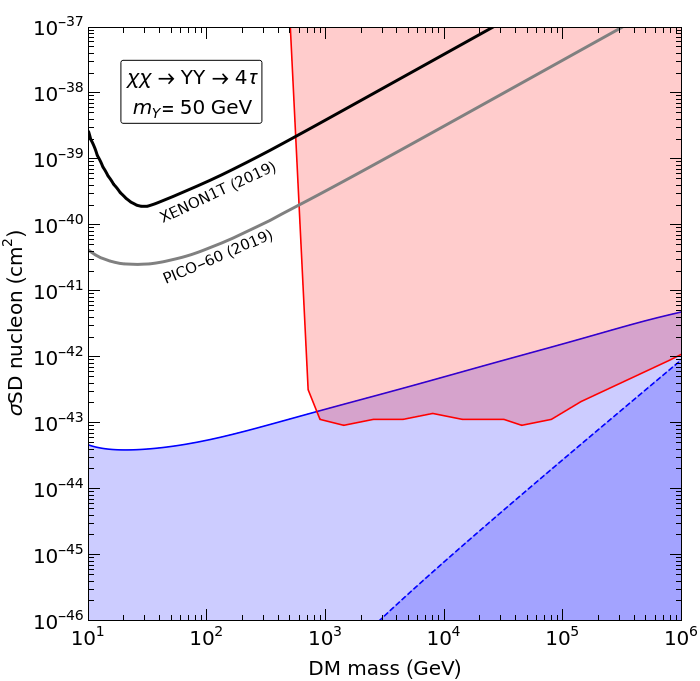} \includegraphics[width=0.45\textwidth]{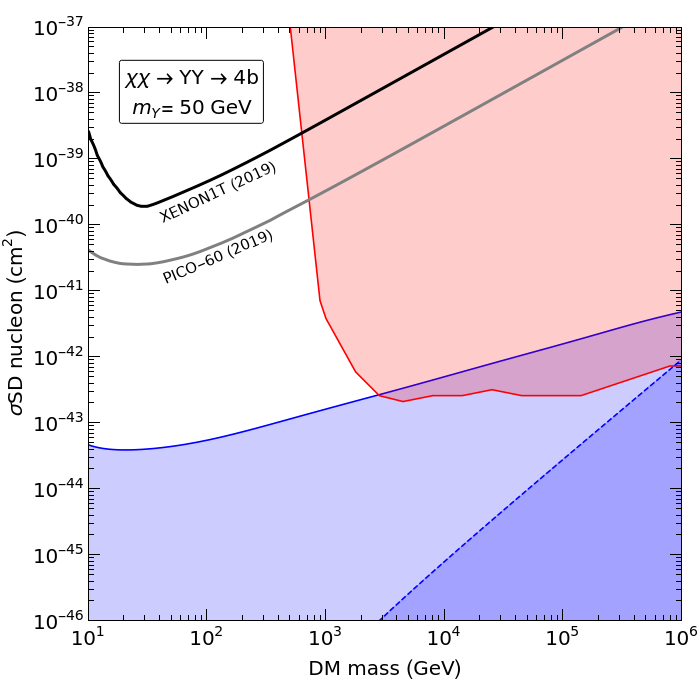} \\
 \caption{SD scattering cross sections excluded by the HAWC gamma ray data (red shaded) for the $\chi\chi \to 4\gamma$ and $4\chi\chi \to4e$ channels, assuming a 1 GeV mediator (upper row) and the $\chi\chi\to4\tau$ and $\chi\chi\to4b$ channels, assuming a 50 GeV mediator (lower row), compared with the XENON1T~\cite{Aprile_2019} and PICO-60~\cite{Amole:2019fdf} direct detection results. The blue shaded regions indicate where the capture-annihilation equilibrium (solid blue line) and thermalisation (dashed blue line) assumptions do not hold.}
    \label{fig:SD4}
\end{figure*}

\begin{figure*}[t]
\includegraphics[width=0.45\textwidth]{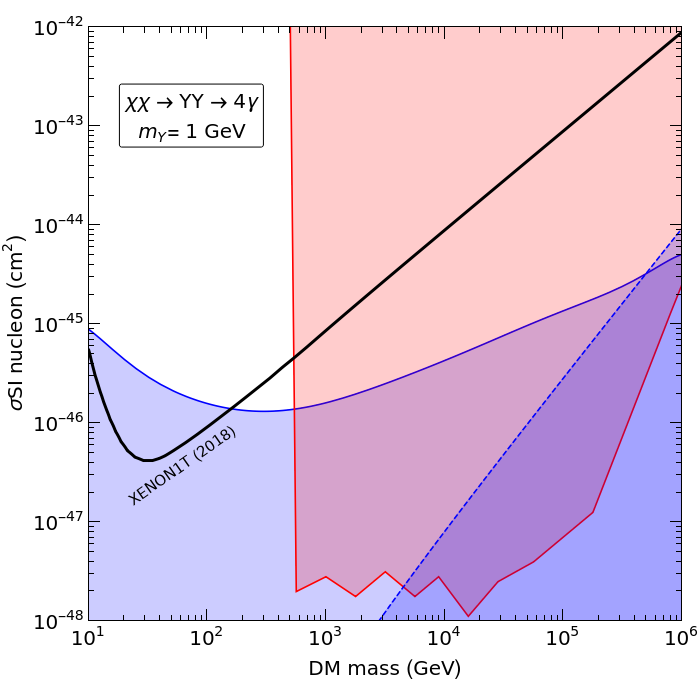} \includegraphics[width=0.45\textwidth]{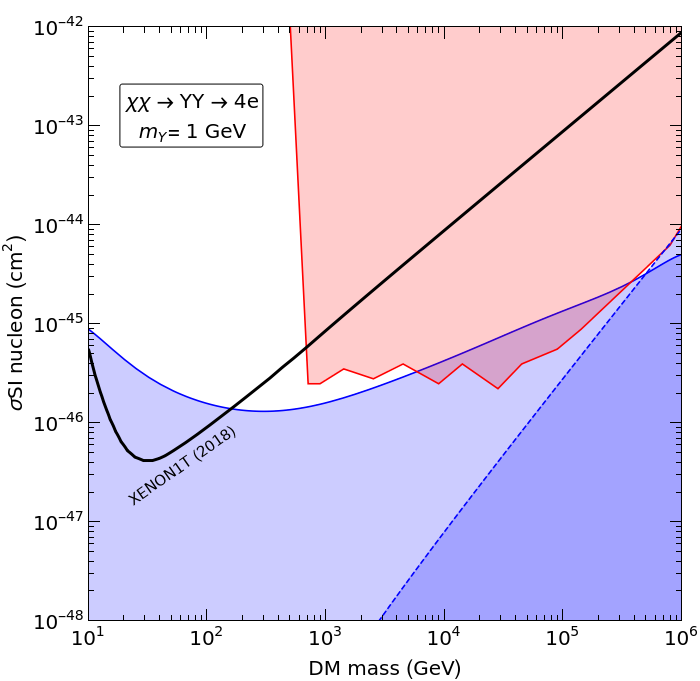} \\
\includegraphics[width=0.45\textwidth]{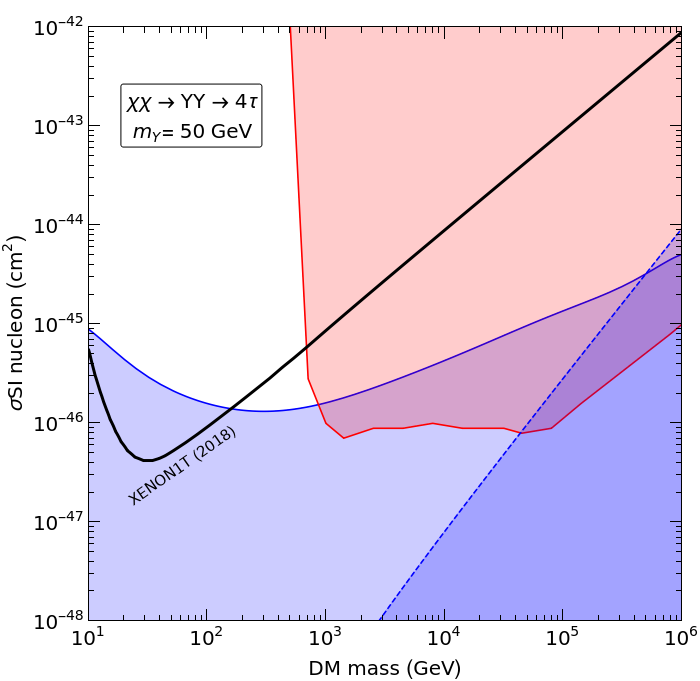} \includegraphics[width=0.45\textwidth]{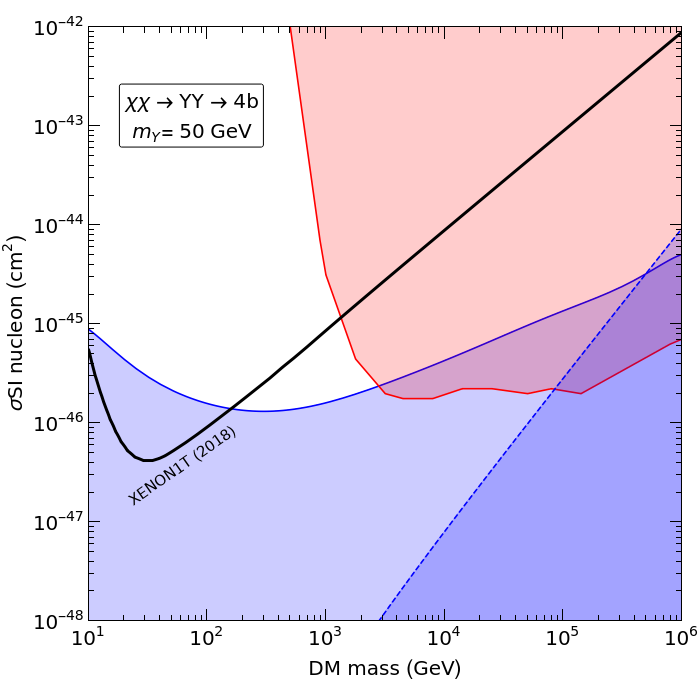} \\
 \caption{SI scattering cross sections excluded by the HAWC gamma ray data (red shaded) for the $\chi\chi \to 4\gamma$ and $4\chi\chi \to4e$ channels, assuming a 1 GeV mediator (upper row) and the $\chi\chi\to4\tau$ and $\chi\chi\to4b$ channels, assuming a 50 GeV mediator (lower low), compared with the XENON1T direct detection results~\cite{Aprile_2018} (black). The blue shaded regions indicate where the capture-annihilation equilibrium (solid blue line) and thermalisation (dashed blue line) assumptions do not hold.}
    \label{fig:SI4}
\end{figure*}

\section{Results} 
\label{sec:results}

\subsection{Model-independent results}
In Figure \ref{fig:SD4} we show the SD cross section excluded by the HAWC gamma ray data for various mediator decay channels, in agreement with~\cite{Albert:2018jwh, Mazziotta:2020foa}.
The shaded blue region indicates the parameters for which capture-annihilation is not in equilibrium. This region was calculated by setting the thermally averaged cross section to the standard freeze-out value,  $\langle \sigma_\text{ann} v \rangle = 2.2 \times 10^{-26}$cm$^3$/s, and agrees with the results of~\cite{Baum_2017}. Note that our assumptions are conservative: if the annihilation cross section were enhanced by the Sommerfeld effect, as occurs naturally in models with light mediators, capture-annihilation equilibrium would be achieved for even smaller scattering cross sections. We also indicate the region in which the DM does not thermalise within the lifetime of the Sun~\cite{Peter:2009mk}. For both the non-equilibrium and non-thermalisation regions, the gamma ray constraints are not applicable because the annihilations will be significantly suppressed.

We see that the solar gamma ray technique provides an extremely sensitive probe of the DM-nucleon scattering cross section. It exceeds the sensitivity of direct detection experiments by many orders of magnitude, for all decay channels, across the entire DM mass range considered.  This is particularly so for a large DM mass, where direct detection constraints become less sensitive.
Moreover, the constraints extend to large masses, where traditional solar DM searches (of the type that search for annihilation directly to neutrinos) lose sensitivity due to the attenuation of high energy neutrinos in the Sun~\cite{Bell:2011sn}.

\begin{figure}[t]
\includegraphics[width=0.45\textwidth]{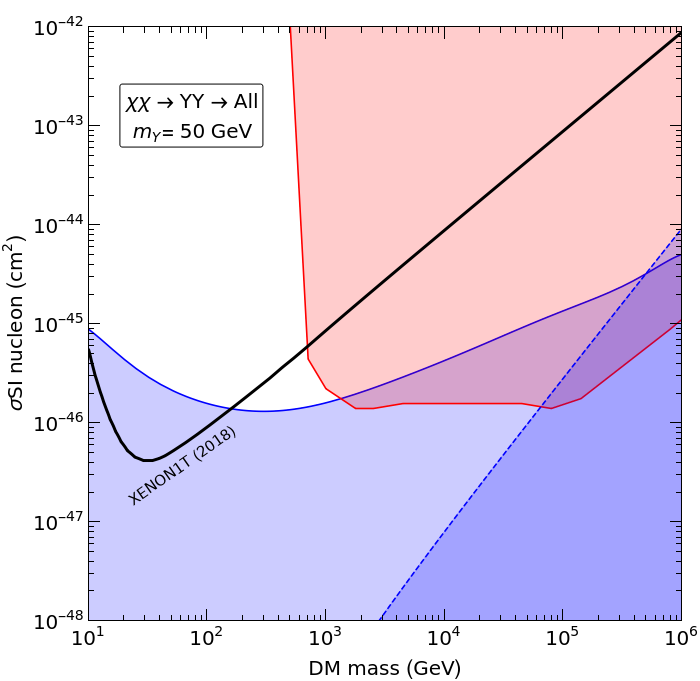}
 \caption{
 SI scattering cross sections excluded by the HAWC gamma ray data (red shaded) for the case of a 50 GeV mediator which decays to all kinematically allowed charged fermion final states, compared with the XENON1T direct detection results~\cite{Aprile_2018} (black). The blue shaded regions indicate where the capture-annihilation equilibrium (solid blue line) and thermalisation (dashed blue line) assumptions do not hold.}
    \label{fig:SI}
\end{figure}

In Fig.~\ref{fig:SI4} and Fig.~\ref{fig:SI} we present similar results for an SI interaction. As with the SD case, we see that the solar gamma ray technique significantly exceeds the sensitivity of direct detection. This comes, perhaps, as a surprise, as it is well known that non--momentum-suppressed and coherently enhanced SI scattering is subject to very strong constraints from direct detection experiments. As such, the solar capture analyses performed to date have focused on SD interactions, making the implicit assumption that the existing direct detection bounds on SI scattering are strong enough to render solar limits irrelevant. However, we see that this is not true. At large DM mass, where the direct detection constraints lose sensitivity, the solar gamma ray technique can probe a sizeable region of previously unconstrained parameter space.

We emphasise that the model-independent results of Figs.~\ref{fig:SD4}, \ref{fig:SI4} and \ref{fig:SI} make a number of simplifying assumptions.  Chief among these are that the DM capture cross section and mediator lifetime are independent parameters and, moreover, that the mediator lifetime is such that $\gamma c\tau$ lies in the range from $R_\odot$ to $D_\oplus$.
Deviating from these assumptions would suppress the gamma ray flux and hence reduce the strength of the cross section limits. This might generically be expected for models in which the scattering cross section and mediator lifetime are controlled by the same set of underlying parameters.
On the other hand, we might expect the non-equilibrium region to be smaller in a realistic model, given the possibility for Sommerfeld enhancement to increase the annihilation rate.  The non-thermalisation region, 
however, will be unchanged.

\subsection{Dark Photon model}

\begin{figure*}[t]
    \includegraphics[width=0.9\columnwidth]{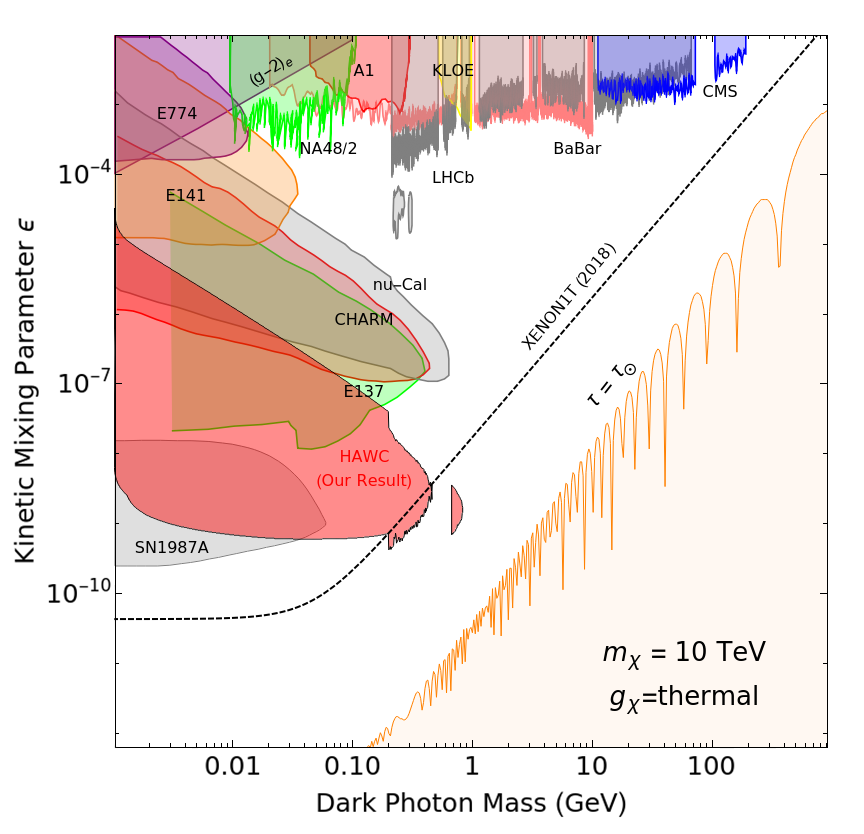}  \includegraphics[width=0.9\columnwidth]{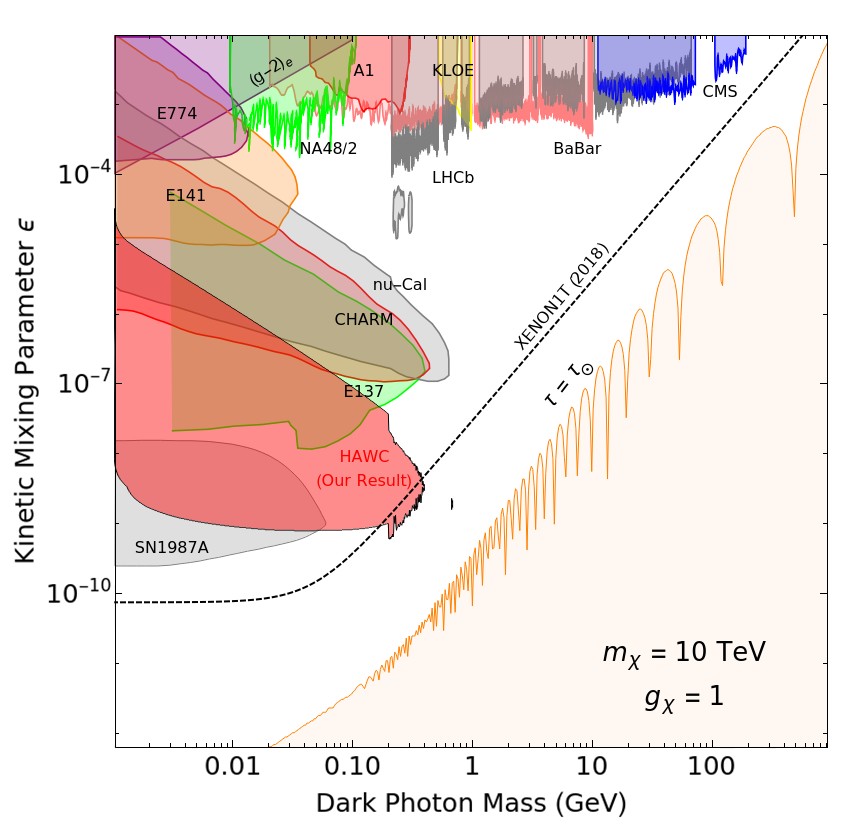} \\
    \includegraphics[width=0.9\columnwidth]{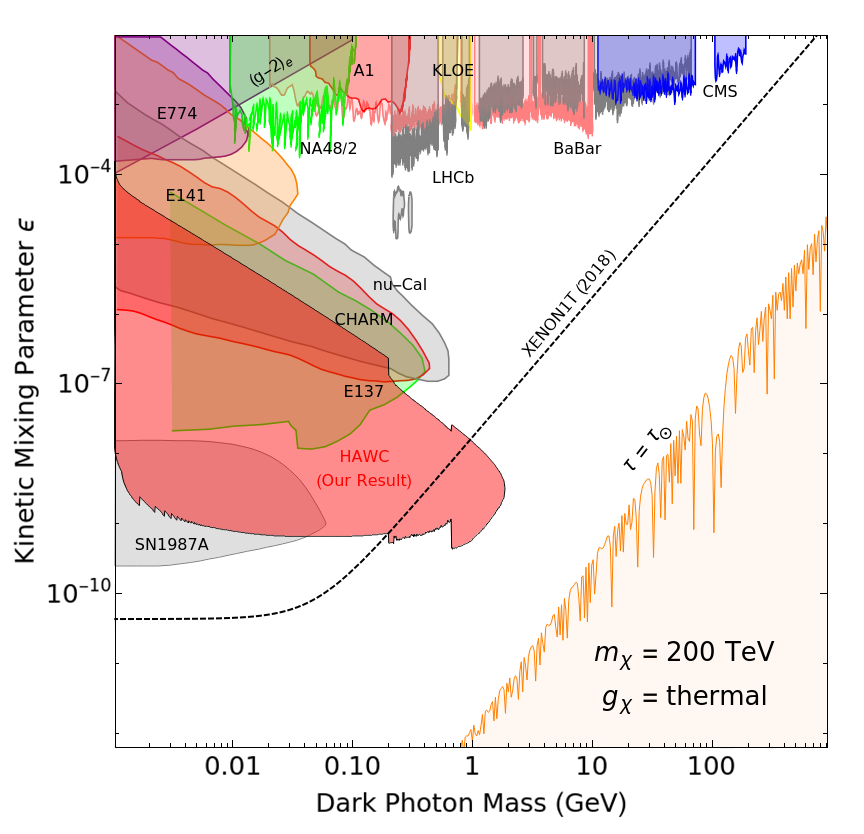} \includegraphics[width=0.9\columnwidth]{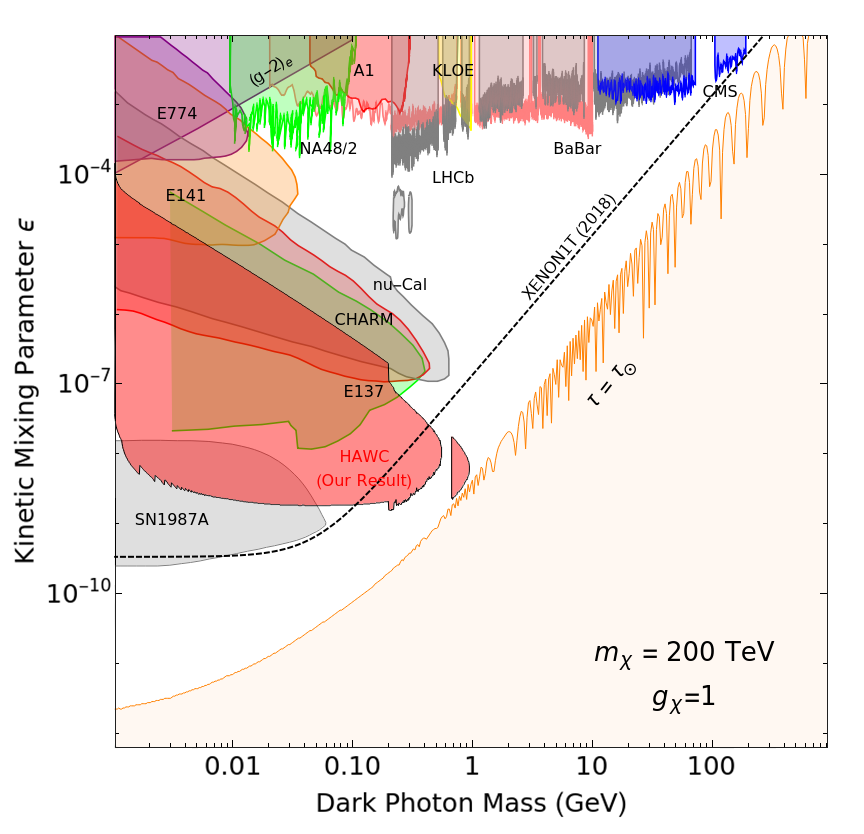} \\
    \caption{Exclusion plot for dark photon model, with DM mass $m_\chi = 10$ TeV (upper row) and $200$ TeV (lower row), and $g_\chi = 1$ (right) or set by thermal freeze-out (left).  Our HAWC gamma ray constraint is shown in red, together with a compilation of current limits taken from~\cite{Fabbrichesi:2020wbt}, using constraints from A1~\cite{Merkel:2014avp}, LHCb~\cite{Aaij:2019bvg}, CMS~\cite{Sirunyan:2018dsf}, BABAR~\cite{Lees:2014xha}, KLOE~\cite{Anastasi:2016ktq,Archilli:2011zc,Babusci:2012cr},  NA48/2~\cite{Batley:2015lha}, E774~\cite{Bross:1989mp}, E141~\cite{Riordan:1987aw}, E137~\cite{Batell:2014mga,Bjorken:1988as,Marsicano:2018krp}, $\nu$-Cal~\cite{Blumlein:2011mv,Blumlein:2013cua}, CHARM~\cite{Gninenko:2012eq}, SN1987A~\cite{Mahoney:2017jqk}. The XENON1T direct detection constraint~\cite{Aprile_2018} assumes a reference momentum of $q_0 = 50$ MeV. The orange region in the lower right of each plot indicates where $\tau_\text{equil} < \tau_\odot$, i.e., where capture-annihilation equilibrium is not achieved.}
    \label{fig:DP4}
\end{figure*}

We now turn to the dark photon model. In Fig.~\ref{fig:DP4}, we show the parameter space constrained by the HAWC gamma ray data, in the $\varepsilon$-$m_{A'}$ plane. Unlike the model-independent scenario, we cannot directly translate this to a $\sigma$-$m_\chi$ constraint due to the energy dependence of the DM-nucleus cross section. Instead, in order to make an approximate comparison with direct detection experiments, we plot the XENON1T results with an assumed momentum transfer of $50$ MeV. 

Our constraints are strongest for larger DM masses, and so we show results for both $m_\chi = 10$ TeV and $200$ TeV. We opt to have $g_\chi$ set either by thermal freezeout -- which requires $\alpha_\chi = 0.024~m_\chi /$TeV, and is in tension with perturbativity at large DM mass -- or we set $g_\chi = 1$ and remain agnostic as to the method of production of the DM relic density.
We can see that the HAWC constraint, shown by the shaded red region, is competitive with the current direct detection constraint from XENON1T, in some regions of parameter space. It is worth noting that both the HAWC and XENON1T constraints scale with $\alpha_\chi$ -- a  smaller coupling implies both a  weaker constraint from HAWC, and a weaker direct detection bound to compete with.

The shaded orange regions in Fig.~\ref{fig:DP4} indicates the parameter space where capture-annihilation equilibrium is not obtained, i.e., where the time scale $\tau_\text{equil} $ is greater than the age of the Sun. The oscillatory features of the equilibrium lines are not a numerical artifact or error; they are due to the resonances of the Sommerfeld enhancement.

The discrete jumps in the HAWC-excluded region around $\sim 200$ and $650$ MeV are due to the kinematic opening of decays to mesons such as pions and $\eta$-mesons, which provide a strong detection signal as they have very discrete, sharply peaked photon spectra compared to bremsstrahlung. 
One might notice small artefacts in the bottom left of the HAWC-excluded region. This, and indeed the curving upward in that region, is a result of the first order approximation of the bremsstrahlung spectrum in Eq.~\eqref{eq:brehm} failing as higher order $m_{A'}/m_e$ effects come into play. In fact we expect the constraint here to level off as the dark photon mass becomes small. In any case, this region of dark photon parameter space is already excluded by SN1987A constraints.

Let us now compare the dark photon results presented in Fig.~\ref{fig:DP4} with the model-independent scenario of Fig.~\ref{fig:SI}. We see that the the HAWC-excluded region of the dark photon model surpasses the XENONIT constraint by a modest amount, with a sensitivity that is lower than might have been expected based on the very strong model-independent results. The reason for this is that the kinetic mixing parameter $\varepsilon$ controls not only the scattering cross section, but also the mediator decay length.
Including this dependence on the decay length in a self-consistent way determines the shape of the red shaded HAWC-excluded regions of Fig~\ref{fig:DP4}. The gamma ray constraints become weak if the dark photons decay too fast, before escaping the Sun (large values of $\varepsilon$, above the excluded region) or if they decay too slowly, beyond the Earth (small values of $\varepsilon$, below the excluded region).

\section{Conclusion}
\label{sec:conclusion}

We have shown that if dark matter annihilates to long-lived mediators, solar gamma ray measurements can be used to place very sensitive constraints on the DM-nucleon scattering cross section that are orders of magnitude more powerful than direct detection experiments.  This is true for both SD scattering, for which our calculations reproduce existing results in the literature, and for SI scattering, which has not previously been considered.

Adopting a simple model-independent approach, in which the mediator is taken to decay between the Sun and the Earth, we demonstrated that the HAWC gamma ray limits exceed the sensitivity of the PICO-60 and XENON1T direct detection experiments for SD and SI scattering, respectively, across the whole dark matter mass range considered. This is especially so at large DM mass, where direct detection experiments lose sensitivity.  

We also made conservative estimates of the parameters for which the assumptions of capture-annihilation equilibrium and dark matter thermalisation in the Sun break down. The solar gamma ray constraints are not applicable in the non-equilibrium and non-thermalisation regions, as the annihilation rate would be significantly suppressed. However, these effects are only applicable for very large DM mass and very small DM cross section. And, in the case of capture-annihilation equilibrium, would be alleviated by the possible presence of Sommerfeld enhancement. Hence the HAWC data still exclude a very large region of previously unconstrained parameter space.  

Finally, we considered a realistic model in which DM annihilates to dark photons, which couple to SM fermions via kinetic mixing of the dark and visible photons.  In this model, the dark photons mediate SI DM-nucleon scattering. The sensitivity of the solar gamma ray technique is reduced, compared to the model-independent analysis, due to the fact that the kinetic mixing parameter controls both the capture rate and the mediator decay length. However, we find the solar gamma ray constraints to be strong and complementary with other bounds on dark photon models, such as those arising from supernova and beam dump experiments, enabling us to exclude a previously unconstrained region of dark photon parameter space.

\medskip

\section*{Acknowledgements}
NFB was supported, in part, by the Australian Research Council and IWS by the Commonwealth of Australia. JBD acknowledges support from the U.S. National Science Foundation under Grant No. NSF PHY-1820801. We acknowledge Adam Green and Flip Tanedo for useful correspondence regarding capture rate calculations, and thank John Beacom, Rebecca Leane and Clarisse Thomas for helpful discussions.


\bibliography{Bibliography} 

\end{document}